\documentclass[aps,prb,twocolumn,longbibliography]{revtex4}

\usepackage{amssymb}
\usepackage{amsfonts}
\usepackage{color}
\usepackage{dsfont}
\usepackage[normalem]{ulem}
\usepackage{dcolumn}
\usepackage{mathtools}
\usepackage{siunitx}
\usepackage[usenames,dvipsnames]{xcolor}
\usepackage[colorlinks=true,linkcolor=blue,citecolor=blue,urlcolor=blue]{hyperref}
\usepackage{graphicx}
\usepackage{blindtext}
\usepackage{outlines}
\usepackage{enumitem}
\usepackage{soul}
\setlist[enumerate,2]{label=\roman*)}
\setlist[enumerate,3]{label=\alph*)}
\usepackage{multirow}

\newcommand{\ud}{\mathrm{d}}

\newcommand{\VEC}[1]{\mathbf{#1}}

\newcommand{\dd}{\ensuremath{\text{d}}}

\newcolumntype{d}[1]{D{.}{.}{#1}}
\definecolor{orange}{rgb}{1,0.5,0}
\definecolor{darkgreen}{RGB}{0,100,0}

\DeclareSIUnit\rydberg{Ry}
\ExplSyntaxOn
\NewDocumentCommand{\au}{m}
 {
 \SI{#1}{{a.u.}}
 \peek_charcode_remove:NT .
  {
  \mode_if_math:F { \spacefactor\sfcode`\.\scan_stop: }
  }
 }
\ExplSyntaxOff

\makeatother

\begin{document}

\title{Spontaneous and externally driven quantum spin fluctuations of 
3$d$ and 4$d$ single atoms adsorbed on graphene}
\author{Siham Sadki$^{1,2}$}
\author{Filipe Souza Mendes Guimar\~aes$^{2}$}
\author{Juba Bouaziz$^{2}$}
\author{Julen Iba\~{n}ez-Azpiroz$^{3}$}
\author{Lalla Btissam Drissi$^{1,2,4}$}
\author{Samir Lounis$^{2}$}\email{s.lounis@fz-juelich.de}

\affiliation{$^{1}$ LPHE, Modeling \& Simulations, Faculty of Science, Mohammed V University in Rabat, Rabat, Morocco}
\affiliation{$^{2}$ Peter Gr\"{u}nberg Institut and Institute for Advanced Simulation,
	Forschungszetrum J\"{u}lich and JARA, J\"{u}lich, Germany}	
\affiliation{$^{3}$ Materials Physics Center, CSIC-UPV/EHU, 20018 Donostia-San Sebasti\'{a}n, Spain}
\affiliation{$^{4}$ CPM, Centre of Physics and Mathematics, Faculty of Science,
	Mohammed V University in Rabat, Rabat, Morocco}


\begin{abstract}
At the heart of current information nanotechnology lies the search for ideal platforms hosting the smallest possible magnets, i.e. single atoms with magnetic moments pointing out-of-plane, as requested in a binary-type of memory. 
For this purpose, a 2D material such as graphene would be an ideal substrate thanks to its intrinsic low electron and phonon densities, as well as its 6-fold symmetry. 
Here we investigate, from first-principles, a fundamental mechanism detrimental for the magnetic stability: the zero-point spin-fluctuations modifying the effective energy landscape perceived by the local spin moments of 3$d$ and 4$d$ transition metal atoms deposited on a free standing graphene. 
Utilizing time-dependent density functional theory and by virtue of the fluctuation-dissipation theorem, these spontaneous quantum fluctuations are found to be negligible for most of the 3$d$ elements, in strong contrast to the 4$d$ atoms.
Surprisingly, we find that such fluctuations can promote the magnetic stability by switching the easy direction of the magnetic moment of Tc from being initially in-plane to out-of-plane. 
The adatom-graphene complex gives rise to impurity states settling in some cases the magnetocrystalline anisotropy energy --- the quantity that defines the energy barrier protecting the magnetic moments and, consequently, the spin-excitation behavior detectable with inelastic scanning tunneling spectroscopy.
A detailed analysis is provided on the impact of electron-hole excitations, damping and lifetime of the spin-excitations on the dynamical behavior of the adsorbed magnetic moments on graphene.
\end{abstract}

\maketitle

\section{Introduction}

In the present days, digital information is stored in binary units.
This is achieved by using building blocks that present bistability, with the two states separated by a reasonable energy barrier.
The utmost miniaturization of these elements in the context of magnetic information storage is a single-atom memory based on the orientation of the atomic magnetic moment.
Of great interest are, thus, magnetic atoms displaying bistability when the rotational symmetry is broken, as, for example, after adsorption on a substrate.
The adatoms may then manifest a magnetic anisotropy favoring the out-of-plane orientation of the atomic moment, i.e., perpendicular magnetic anisotropy (PMA), as opposed to an anisotropy that favours the moment to lay in-plane, where it can rotate 360$^\circ$ with ideally no energy dissipation. 
This drives the current race for the discovery and engineering of platforms enhancing the PMA and magnetic bistability of adatoms for information technology~\cite{Donati2016,Khajetoorians2016,Hermenau2017,Willke2019}.

For technological applications, it is also important that the substrate possesses advantageous electrical properties~\cite{ruckh_1996,azpiroz_2013,koster_2012}.
Graphene is appealing thanks to its fascinating physical charcteristics~\cite{Novoselov2004,Novoselov2005,Neto2009,Kim:2009ed,Han:2014dc} such as high carriers mobility~\cite{Maassen:2010kg} and excellent conductivity~\cite{Wang:2010ba}. 
This material also exhibits very long spin relaxation times (in the range of \SI{}{\micro\second}), and relaxation lengths (of the order of \SI{100}{\micro\meter})~\cite{Kane:2005hl,HuertasHernando:2006ha}.
All in all, graphene is one of the most promising candidates for spin information transport since it can be used as a spin-conserver system that can transmit spin-information with high fidelity~\cite{HuertasHernando:2006ha,Pesin:2012hj}.

Various theoretical investigations demonstrated that transition elements embedded in graphene can develop finite magnetic moments with sizeable magnetocrystalline anisotropy energy (MAE), see e.g. Refs.~\citenum{Mao:2008kb,Sargolzaei:2011dm,Wehling:2011dw,Hu:2014bt}, which motivated several studies devoted to transport properties.
For example, heavy 4$d$ and $5d$ adatoms were theoretically prospected to realize the quantum spin Hall state in graphene~\cite{Weeks:2011cu,dosSantos:2018gq}.
In addition, a giant 2D topological insulating gap was predicted in graphene decorated with heavy adatoms possessing partially filled $d$-shells such as osmium and iridium~\cite{Hu:2012iq}.
Despite their possible impact on the Hall effects, recent predictions indicate that some impurities can be non-detectable electrically if deposited in one of most common adsorption sites of graphene\cite{Duffy:2016gi,Tuovinen:2019ke}. 
When the adsorbed moments are driven to an excited state, as shown from a simple model approach, the resulting dynamical spin-excitations are carried for long distances without dispersion in nanotubes and nanoribbons~\cite{Guimaraes:2010fi}, which can be utilized as spin transistors that can be turned on or off with a purely electric gate~\cite{Guimaraes:2010ha}.

Several ingredients, such as its intrinsic low electron and phonon densities, seem to favor the exploration of graphene as an ideal host for stable magnetic moments. 
However, studies on other substrates, such as metals and insulators, pinpointed several mechanisms challenging the robustness of bistability favored by the PMA, e.g. Refs.~\cite{Miyamachi2013,Huebner2014,Khajetoorians2013,Donati2016,IbanezAzpiroz:2016fa}, hitherto not considered for graphene. 
A major issue identified recently are spontaneous quantum fluctuations present even at absolute zero temperature, the so-called zero-point spin-fluctuations (ZPSF)~\cite{IbanezAzpiroz:2016fa,IbanezAzpiroz:2017dv,IbanezAzpiroz:2018}. 
The magnitude of the latter is intimately related to the spin-excitations properties shaped by the local magnetic moments, the strength of spin-orbit coupling and electron-hole interactions.
	
X-ray magnetic circular dichroism (XMCD) and inelastic scanning tunneling spectroscopy (ISTS) identified several adatoms with MAE of a few \SI{}{\milli\electronvolt} (see e.g. Refs.~\citenum{Gambardella:2003iz,Hirjibehedin2007,Khajetoorians2013,Donati2014,Rau2014,Singha2017}), which surprisingly behave paramagnetically when probed using XMCD or spin-polarized scanning tunneling microscopy (SP-STM). 
This is the case for Co and Fe adatoms, which were found to be paramagnetic, and carrying a magnetic moment with an out-of-plane easy axis when deposited on Graphene/SiC~\cite{Eelbo:2013fta}.
In contrast, Ni seemed to be non-magnetic. 
Recently, measurements based on ISTS~\cite{Donati:2013is} pointed out that single Co adatoms are also paramagnetic on Pt-supported graphene with a MAE of $\SI{8.1}{\milli\electronvolt}$ favouring a PMA. 
So far, the only single adatoms demonstrated to be stable magnetically by showing a hysteresis opening are Ho and Dy atoms deposited on respectively insulating MgO films on Ag(001) surface and graphene on top of Ir(111)~\cite{Donati2016,Baltic2016}.

In this work, we explore by means of first-principles utilizing time-dependent spin-density functional theory (TD-DFT) the magnetic stability and spin-excitations of a set of 3$d$ (Mn, Fe, Co) and 4$d$ (Mo, Tc, Ru) transition metal adatoms deposited on the hollow site, i.e. at the center of the hexagon, of a graphene monolayer. 
On the one hand, we obtain spin-excitation spectra detectable experimentally with ISTS, i.e. spin-fluctuations that can be externally driven, which encode all the information essential to the atomic spin-dynamics such as the MAE, electron-hole excitations and excitation lifetimes. 
These are important to characterize for any manipulation of the state of a magnetic moment. On the other hand, we employ the fluctuation-dissipation theorem~\cite{Kubo:2002dq} to evaluate the ZPSF, which modify the energy landscape of the magnetic adatoms and renormalize the magnitude of the MAE as obtained from simulations based on regular static DFT. 
The 3$d$ atoms experience a relatively slight decrease of their effective MAE, with Fe having the largest PMA. 
Thus graphene offers an ideal platform for weak ZPSF and, by extension, for stable moments. 
The 4$d$ atoms with their moments preferring initially to be in-plane, however, develop large ZPSF decreasing thereby their respective MAE. 
Interestingly, we discovered that the moment of one of the adatoms, Tc, switches its preferable orientation to point out-of-plane. 
In other words, ZPSF counter-intuitively can promote PMA on graphene. 
It is to our knowledge the first element demonstrating such a behavior.

We find that the spin excitational behavior can be deeply impacted by the localized impurity states emerging near the Fermi level as a result of the hybridization of the electronic states of the transition metal adatoms and of graphene.
This is particularly strong in the case of Fe due to its large $d$-resonance at the Fermi energy in the minority channel.
In combination with doping and electric fields, this feature can be used to manipulate the excitations --- both driven and spontaneous --- to achieve a tunable stability that can be used in future magnetic storage devices.

This manuscript is structured as follows: Section~\ref{sec:methods} provides the computational details and methods on which our study is based. 
The results of the simulations and their analysis are displayed in Section~\ref{sec:results}, where we present both static ground state properties and the excitations/fluctuations, and discuss how these quantities are intertwined.
Finally, we conclude in Section~\ref{sec:conclusions} and provide an outlook for future work in this direction. 

\section{Computational methods}
\label{sec:methods}

The ground state of the investigated systems are obtained within density functional theory.
Initially, the structural relaxations of the impurities are determined using the Quantum Espresso simulation package~\cite{Giannozzi:2009hx} within the local spin density approximation (LSDA)~\cite{Perdew:1981dv}.
The obtained distances are listed in Table~\ref{table1}.
We have also obtained similar results using the generalized gradient approximation (GGA). 
The relaxations were done setting the convergence criterion of the force to \au{e-4}, with a periodic $4\times 4$ supercell containing 32 carbon atoms and the impurity. 
An energy cutoff of $\SI{70}{\rydberg}$ was assumed. 
Once the relaxed positions are computed, they are used as inputs for simulations using the Korringa-Kohn-Rostoker (KKR) Green function method~\cite{Papanikolaou:2002hk,Bauer2013}, which are done in two steps: First, we self-consistently determine the ground state of the periodic graphene monolayer within the atomic-sphere approximation including the full charge density. 
Second, the transition metal impurities are embedded into a real space impurity cluster containing $20$ neighbouring carbon atoms.
This method does not suffer from the problem of having periodically-repeated impurities that couple to each other~\cite{Venezuela:2009bg}.
The spin-orbit interaction is included self-consistently.

Starting from the obtained ground state, we then study the spin excitations described by the dynamical spin response of the 3$d$ and 4$d$ impurities deposited on graphene using  TD-DFT~\cite{Lounis2010,Lounis2011,Lounis:2015ho, dosSantosDias:2015bh}. 
The central quantity in this approach is the dynamical magnetic susceptibility, which is defined as the spin magnetization response to an oscillatory magnetic field with frequency $\omega$, $\delta \VEC{b}(\VEC{r},\omega)$, as
\begin{equation}
\delta m_{\alpha}(\VEC{r},\omega) = \sum_{j}\int\!\dd\VEC{r}^{\,\prime}\,
\chi_{\alpha\beta}(\VEC{r},\VEC{r}^{\,\prime},\omega)\,\delta 
b_{\beta}(\VEC{r}^{\,\prime},\omega)\quad.
\label{liner_response}
\end{equation}
where $\alpha,\beta$ represent cartesian components. 
We focus only on the transversal spin excitations~\cite{dosSantosDias:2015bh,IbanezAzpiroz:2016fa}, which are encoded in the spin-flip dynamical magnetic susceptibility, $\chi^{+-}(\omega)$, since they correspond to the spectra measured within ISTS~\cite{Lounis2010,Khajetoorians2011}.
It can be computed from the non-interacting Kohn-Sham susceptibility $\chi^{+-}_\text{KS}(\omega)$ after solving the Dyson-like equation
\begin{equation}
\chi^{+-}(\omega)=\left[ 1-\chi_\text{KS}^{+-}(\omega)K_\text{xc}\right]^{-1}\chi^{+-}_\text{KS}(\omega)\quad.
\label{full}
\end{equation}%
$K_\text{xc}$ is the exchange-correlation kernel defined in the adiabatic local spin-density approximation~\cite{dosSantosDias:2015bh}, i.e., frequency independent and local in space.
Furthermore, a more intuitive understanding of the spin dynamics can be obtained by mapping the susceptibility obtained from first-principles into the one from a generalized phenomenological Landau-Lifshitz-Gilbert (LLG) model~\cite{Gilbert:2004gx}, where the equation of motion of the magnetic moment ${\VEC{M}} = \int_{V_\text{cell}} \ud\VEC{r}\, \VEC{m}(\VEC{r})$ ($V_\text{cell}$ being the volume of the magnetic unit) reads
\begin{equation}
\frac{\ud\VEC{M}}{\ud t}=-\gamma\,(\VEC{M}\times 
\VEC{B}_\text{eff})+\alpha\,\frac{\VEC{M}}{{M}}\times \frac{\ud\VEC{M}}{\ud t} 
\quad.
\label{LLG_eq}
\end{equation}
The first term on the right-hand side is proportional to the gyromagnetic ratio $\gamma$ and represents the torque that drives the magnetization into precession motion, the second one accounts for the damping through the Gilbert damping parameter $\alpha$~\cite{Gilbert:2004gx,Guimaraes:2019bs}.
The effective magnetic field $\VEC{B}_\text{eff} = \VEC{B}_\text{ext}-\frac{2\mathcal{K}}{M}\VEC{\hat{e}}_{z}$ originates in the external field as well as in the magnetic anisotropy energy, which, for the uniaxial system under consideration, reads
\begin{equation}\label{mae_energy}
\mathcal{E}_{a} = -\mathcal{K}\,\left(\frac{\VEC{M}}{M}\cdot\VEC{\hat{e}}_{z}\right)^2\quad.  
\end{equation}
$\mathcal{K}$ is the magnetic anisotropy constant, which can be extracted from the magnetic susceptibility, $\mathcal{K}_\text{susc}$, or from the band energy differences, $\mathcal{K}_\text{b}$.~\cite{Bouaziz:2019bf}

\section{Results}
\label{sec:results}

In this section, we first study and discuss ground state properties of the 3$d$ and 4$d$ impurities adsorbed on graphene. 
Subsequently, we analyze the spin-excitation spectra and relate it to the physical quantities present in Eq.~\eqref{LLG_eq}. 
Third, we examine the spin-fluctuations and their impact of the magnetic anisotropy barriers of the impurities under consideration.

\begin{figure}
\includegraphics[width=1.0\columnwidth]{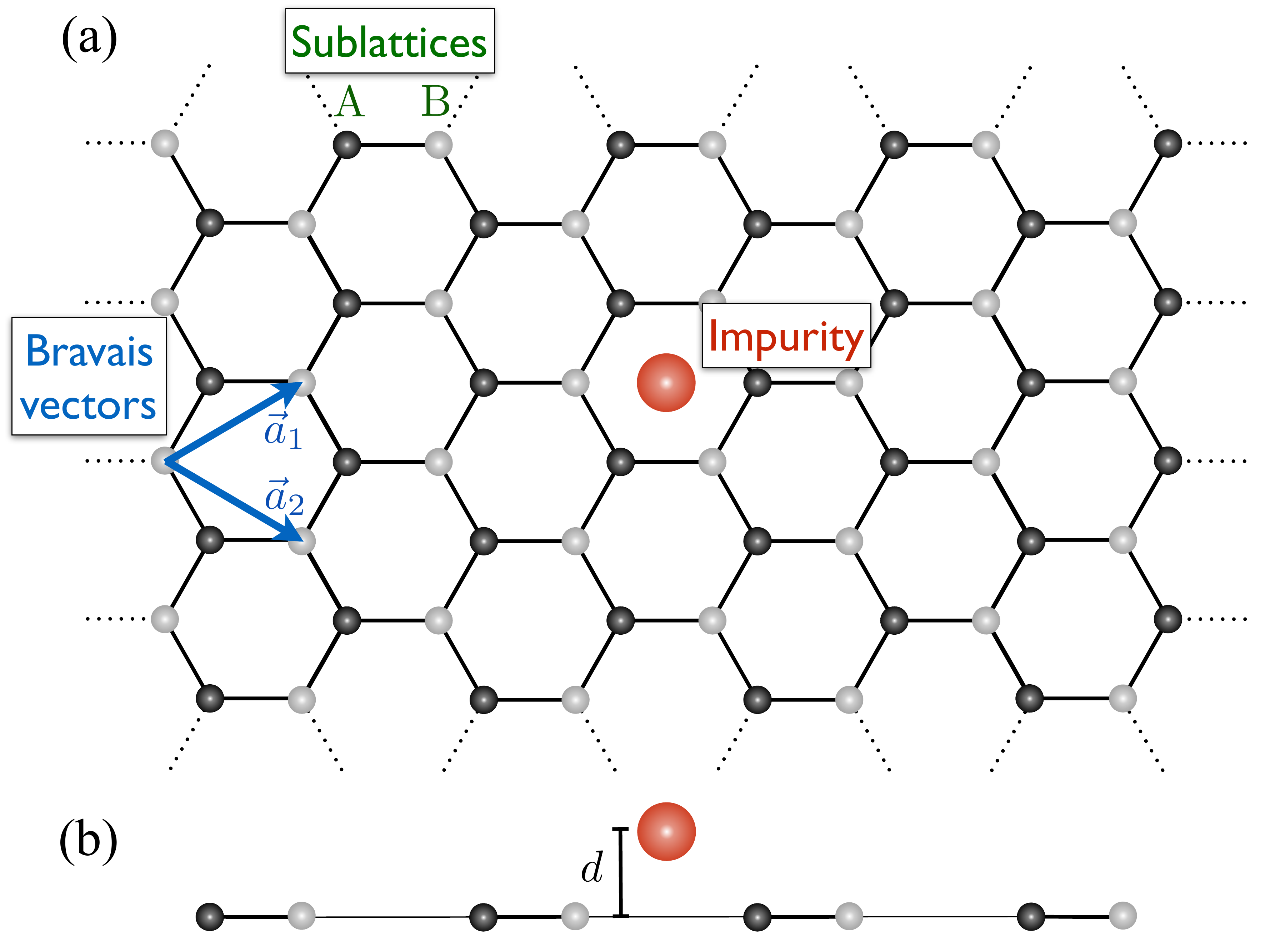}
\caption{a) Top view of the graphene monolayer with an impurity adsorbed on a hollow site. The unit cell of graphene is composed by two sites, one for each sublattice, A and B. The Bravais vectors $\VEC{a}_1$ and $\VEC{a}_2$ are depicted in green.
b) Side view displaying the relaxed distance $d$ between the impurity and the graphene monolayer. 
The values of $d$ are listed in Table~\ref{table1}.}
\label{diagram}
\end{figure}

The geometrical configuration for the adatom adsorbed on graphene hollow-site is depicted in Fig.~\ref{diagram}. 
Depending on their chemical nature, the different impurities relax towards the graphene monolayer at different heights ranging from ${d} = 1.5$ to $\SI{2.9}{\angstrom}$. 
The relaxed positions, $d$, for each impurity are given in Table~\ref{table1}.
The computed values are in good agreement with previous DFT-based calculations (see Refs.~\onlinecite{Wehling:2011dw, Mao:2008kb}). 
Note that, overall, the 3$d$ adatoms tend to relax more in comparison with the 4$d$ elements. 
This is due to the fact that 4$d$ orbitals are more extended spatially than the 3$d$ ones, which affects the effective size of the atoms and, therefore, the resulting atomic relaxation. 

\begin{table}[tbp]
	\centering%
	\begin{tabular}{cccc}
		\hline
		\text{Element} & \text{$d$ (\AA )} &  $M_\text{s}$ ($\mu_\text{B}$) &  $M_\text{o}$ ($\mu_\text{B}$) \\ \hline
		\text{Mn} & 2.3 &   3.57 & 0.01 \\ 
		\text{Fe} & 1.9 &   2.86 & 0.09 \\ 
		\text{Co} & 1.5 &   1.52 & 0.05 \\ \hline 
		\text{Mo} & 2.9 &  1.33 & 0.05 \\ 
		\text{Tc} & 2.2 &   1.05 & 0.11 \\ 
		\text{Ru} & 2.4 &   1.86 & 0.05 \\ \hline 
	\end{tabular}%
	\caption{ Ground state properties for 3$d$ and 4$d$ elements adsorbed on graphene. 
		The height (in $\SI{}{\angstrom}$) is defined as the perpendicular distance between the transition metal elements and the surface of graphene, as shown in Fig.~\ref{diagram}. 
		$M_{s}$ and $M_{o}$ are the spin and orbital magnetic moments, respectively. }
	\label{table1}
\end{table}

\subsection{Ground state properties}
\label{groundstate}

\begin{figure*}
\includegraphics[width=1.0\textwidth]{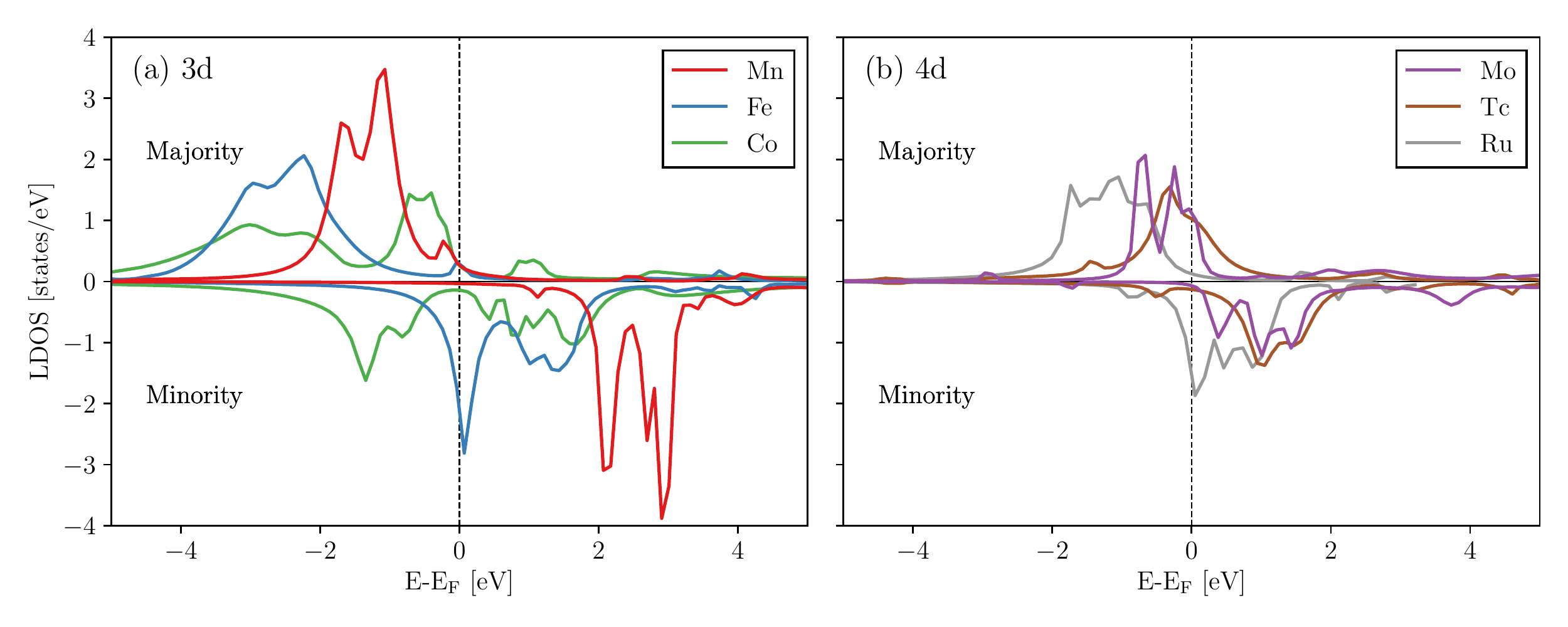}
\caption{Local density of states of (a) 3$d$ and (b) 4$d$ adatoms deposited on the center of the hexagon of a graphene sheet.}
\label{ldos}
\end{figure*}

The electronic states of the adatoms can be visualized in the local density of states (LDOS) shown in Fig.~\ref{ldos}.
We note that apart from the usual peaks expected from the $d$ resonances of the adatoms, localized features, which are nothing else than impurity-states, show up close to the Fermi level.
This is more evident in the case of Mn and Fe, where this state stands out in the majority-spin channel.
Fe presents also a very high density of states in the minority-spin channel, which can also be originating  or have a contribution from the impurity-state.
We note that the LDOS of Co forms a rather wide band resulting from the large hybridization with the electronic states of the graphene sheet induced by the short adatom-substrate distance ($d=\SI{1.5}{\angstrom}$).
In contrast to the 3$d$ elements, the exchange splitting between the majority- and minority-spin bands for the 4$d$ adatoms is smaller, as expected.
This correlates with the spin moments $M_s$ obtained in Table~\ref{table1}.

The 3$d$ elements display a maximum value of $M_s$ at half filling, i.e., for Mn with $M_\text{s} = 3.57\,{\mu}_\text{B}$, and a minimal value for Co, $M_\text{s} = 1.52\,{\mu}_\text{B}$ --- in line with the predictions based on Hund's rules and previous results obtained for similar systems~\cite{Mao:2008kb,Sargolzaei:2011dm}. 
Surprisingly, the orbital moment $M_\text{o}$ for Co is found to be half of the one computed for Fe, in contrast to the larger expected value for isolated atoms.
Such a scenario may occur for the systems under consideration due to the strong hybridization of the electronic states.

The spin moments of the 4$d$ adatoms experience first a decrease when going from Mo (1.33 $\mu_B$) to Tc (1.05 $\mu_B$), followed by an increase for Ru (1.86 $\mu_B$) as a function of the filling of the electronic states. 
This is in line with the relaxations experienced by each of the impurities. 
Indeed, Tc is the closest to the substrate, which enhances hybridization and reduces the magnitude of the spin moment.
Due to their stronger spin-orbit interaction, the values of the orbital moments obtained for the 4$d$ elements are generally larger than the ones found for the 3$d$ adatoms, in contrast to the spin moments. 

Upon rotation of the spin moment, the orbital moment can experience a modification that, under some approximations, can be related to the MAE by the so-called Bruno's formula~\cite{Bruno:1989bz}. 
This expression stipulates that the spin moment prefers to align along the direction maximizing the orbital moment, with the MAE proportional to the anisotropy of the orbital moment $\Delta M_{o} = M_{o}^z-M_{o}^x$ weighted by the strength of the atomic spin-orbit interaction ($z$ and $x$ being the out-of-plane and in-plane directions, respectively).
The values of $\Delta M_{o}$ for the investigated impurities are listed in Table~\ref{table2}, where a positive value indicates a PMA. 
Based on Bruno's formula, Fe, Co and Ru are expected to display a PMA, contrary to Mn, Mo and Tc. 
Furthermore, we expect the MAE of the 4$d$ elements to be larger than those of the 3$d$ ones since the orbital magnetic anisotropy is of the same order of magnitude but the spin-orbit interaction is the largest for the former atoms.

In practice, we compute the MAE employing two distinct methods: band energy differences between two orientations of the spin moment, and linear response theory in the spirit of TD-DFT.
 The former method is often used in the literature, it does not necessarily lead to values of the MAE compatible with the one acting on the spin-excitation spectrum~\cite{Bouaziz:2019bf} discussed in the next section.
If one is addressing an idealized Heisenberg system, where the spin moments do not change upon rotation, both schemes should lead to similar results.

\begin{figure}[tbp]
\includegraphics[width=1.0\columnwidth]{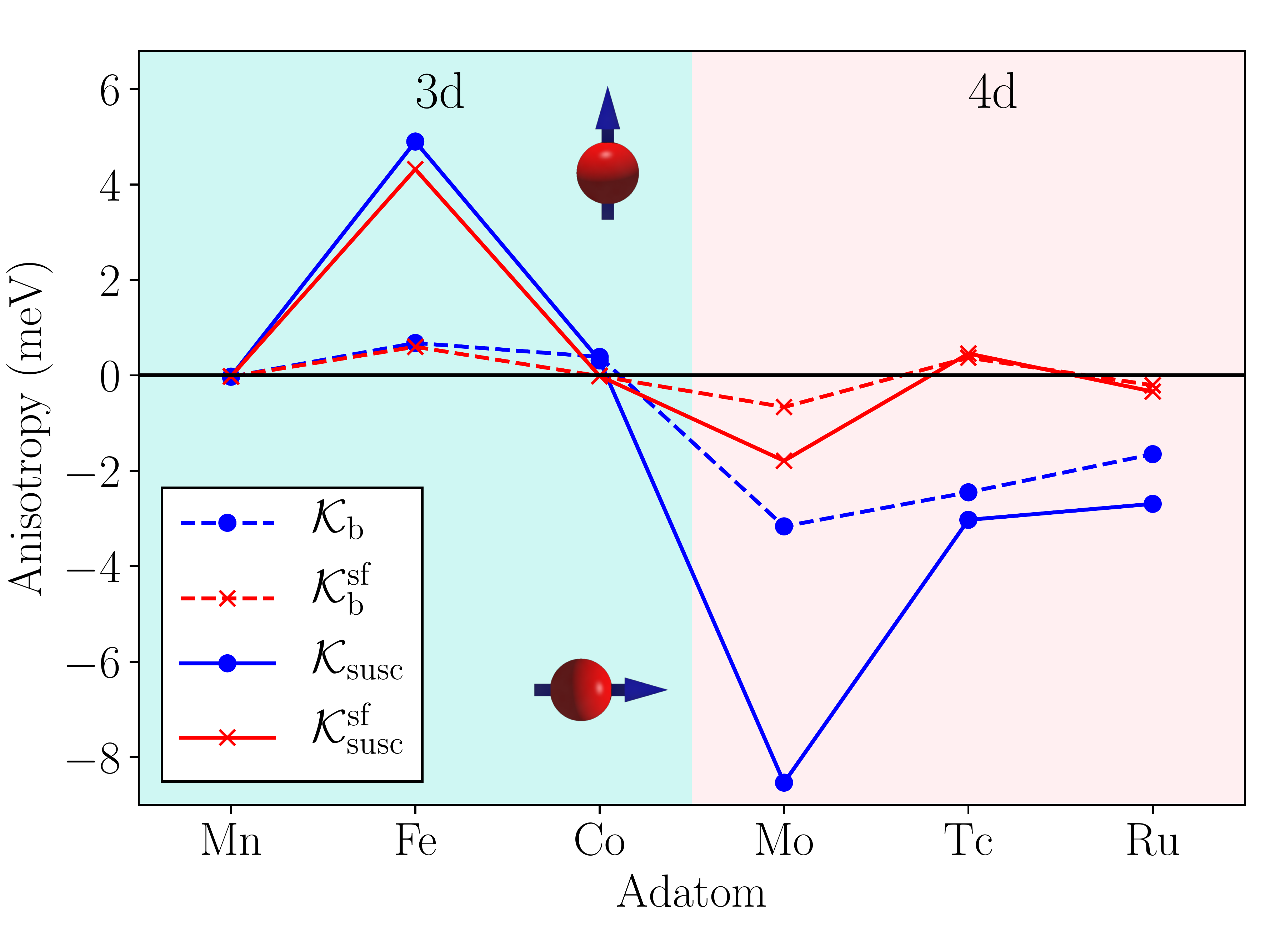}
\caption{ 
Values of the bare MAE for 3$d$ and 4$d$ adatoms deposited on graphene using the magnetic force theorem, $\mathcal{K}_\text{b}$, and magnetic susceptibility, $\mathcal{K}_\text{susc}$, as well as their renormalized values due to zero-point spin fluctuations, $\mathcal{K}^\text{sf}_\text{b}$ and $\mathcal{K}^\text{sf}_\text{susc}$, respectively.
Positive values indicate a preferable out-of-plane orientation of the magnetic moment. 
}
\label{fig33}
\end{figure}

We start by analysing the band energy method.
Within the framework of the magnetic force theorem~\cite{Oswald:1985hs,Liechtenstein:1987br,Daalderop:1990fd}, the MAE can be obtained from Eq.~\eqref{mae_energy} as $\mathcal{K}_\text{b} = \mathcal{E}_\text{x} - 
\mathcal{E}_\text{z}$, where $\mathcal{E}_\alpha$ represents the band energy when the spin points along $\alpha$. 
Note that within our sign convention and similarly to the orbital moment anisotropy, positive and negative $\mathcal{K}_\text{b}$ correspond to preferred out-of-plane and in-plane magnetization directions, respectively. 
The obtained values of the MAE for the different elements are plotted in Fig.~\ref{fig33} and listed in Table~\ref{table2}. 
Among the 3$d$ elements, Fe and Co favor an out-of-plane orientation, in contrast to Mn that points in the plane.
All the 4$d$ impurities display an easy plane anisotropy, according to the band energy differences.
We also notice that the MAE of 4$d$ elements is substantially larger in comparison to the 3$d$ ones, due to their stronger spin-orbit interaction (from which the MAE originates), as expected from Bruno's formula. 

\begin{table*}
\centering%
\begin{tabular}{ccccccccccccccc}
\hline
\text{Element} & $\Delta M_{o} (\mu_\text{B})$ & $\mathcal{K}_\text{b}$ & $\mathcal{K}^\text{sf}_\text{b}$ 
& $\mathcal{K}_\text{susc}$ &  $\mathcal{K}^\text{sf}_\text{susc}$ & $\omega_\text{max}(\SI{}{\milli\electronvolt})$ 
& $\alpha$ &  $\tau(\SI{}{\pico\second})$ & $\frac{\xi_\perp }{M_\text{s}}(\%)$ \\ \hline
\text{Mn} & -0.012 & -0.029 &  -0.025 &  -0.026 & -0.022 &  0.022 & 0.014 & 14038 & 21 \\ 
\text{Fe} & 0.047 &  0.679 &  0.599 &  4.900 &  4.320 &  0.220 & 53.41 & 0.201 & 21 \\ 
\text{Co} & 0.036 &  0.386 &  -0.017 &  0.313 & -0.013 &  0.644 & 0.033 & 191.8 & 108 \\ \hline
\text{Mo} & -0.067 & -3.165 &  -0.664 &  -8.536 & -1.791 &  9.399 & 0.012 & 34.14 & 75 \\ 
\text{Tc} & -0.078 & -2.451 &  0.370 &  -3.028 &  0.457 &  5.827 & 0.409 & 1.672 & 130 \\ 
\text{Ru} & 0.006 & -1.652 &  -0.208 &  -2.692 & -0.339 &  2.708 & 0.163 & 9.322 & 84 \\ \hline
\end{tabular}
\caption{Magnetic properties of 3$d$ and 4$d$ elements adsorbed on a graphene monolayer. 
$\Delta M_{o}$ accounts for the orbital moment anisotropy. 
The constant $\mathcal{K}_\text{b}$ denotes the MAE calculated using the magnetic force theorem, while $\mathcal{K}^\text{sf}_\text{b}$ represents its value renormalized by the spin-fluctuations.
$\mathcal{K}_\text{susc}$ and $\mathcal{K}^\text{sf}_\text{susc}$ designate the same quantities, but obtained from the magnetic susceptibility. 
All the anisotropy constants are given in \SI{}{\milli\electronvolt}. 
$\omega_\text{max}$ is the resonance frequency of the spin excitation, $\alpha$ is dimensionless Gilbert damping parameter, and $\tau$ represents the lifetime in picoseconds.
We also list the ratio between the zero-point fluctuation amplitude and the spin magnetic moment $\frac{\xi}{M_\text{s}}$ for each impurity.
}
\label{table2}
\end{table*}

The trends (sign and magnitude) of $\mathcal{K}_\text{b}$ obtained for the 3$d$ impurities are in general in good agreement with the predictions of Bruno's formula. 
For example, Mn has nearly half-filled $d$-orbitals, presenting the lowest orbital moment among the investigated adatoms, $M_\text{o} = 0.01\mu_\text{B}$ (see Table~\ref{table1}).
Accordingly, it also presents a weak orbital magnetic anisotropy and, consequently, a very small in-plane MAE, as listed in Table~\ref{table2}. 
We note that Bruno's formula fails to describe the easy axis of Ru.

In addition to using band-energy differences, the MAE can alternatively be obtained from the static transverse magnetic susceptibility, calculated within linear response theory, as $\mathcal{K}_\text{susc} = M_s^2\operatorname{Re}[\chi^{-+}(\omega=0)]^{-1}$
~\cite{Guimaraes:2019bs,Bouaziz:2019bf}.
The values obtained using this approach are also given in Table~\ref{table2} and illustrated in Fig.~\ref{fig33}.
The two methods disagree when sharp features are present near the Fermi energy in the density of states.
This causes large changes in the electronic structure when rotating the magnetic moment, which invalidates the use of the magnetic force theorem~\cite{Bouaziz:2019bf}.
Fe and Mo are the most affected elements (see Fig.~\ref{ldos}), with values of $\mathcal{K}_\text{susc}$ reaching $\SI{4.9}{\milli\electronvolt}$ and $\SI{-8.5}{\milli\electronvolt}$, respectively.

Finally, we point out that our results are in line with the experimental and theoretical results reported in Ref.~\onlinecite{Eelbo:2013fta} for Fe and Co impurities adsorbed on a monolayer graphene on SiC(0001).

\subsection{Spin-excitations spectra}

\begin{figure*}[tbp]
\includegraphics[width=1.0\textwidth]{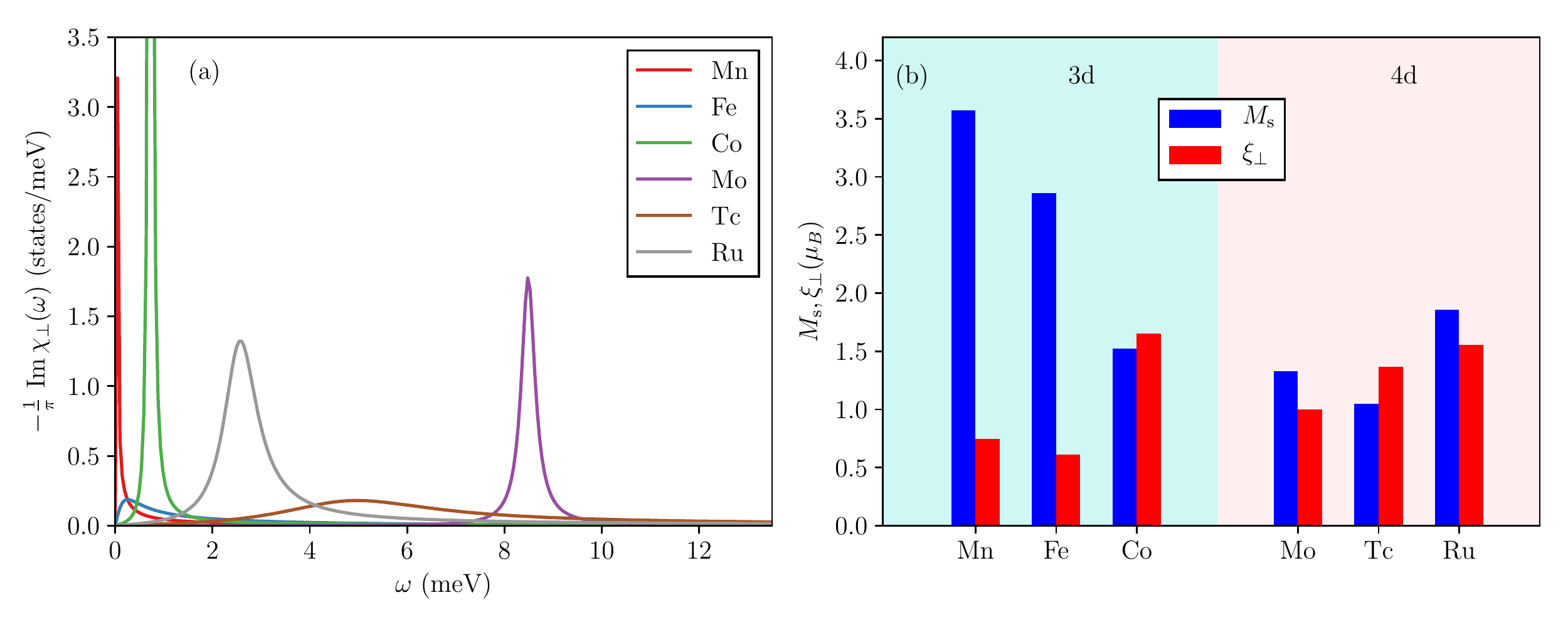}
\caption{(a) Density of transverse spin excitations calculated for 3$d$ and 4$d$ elements adsorbed on graphene. (b) The calculated zero-point spin-fluctuations compared to the spin magnetic moment.}
\label{spin-excitation}
\end{figure*}

In this section, we analyze the transverse magnetic susceptibility, which describes the excitations driven by time-dependent external magnetic fields as given, in Eq.~\eqref{liner_response}.
The density of transverse spin-excitations is obtained from the imaginary part of
\begin{equation}
\chi_{\bot}\left(\omega\right) = \left[\chi_{+-}(\omega)+\chi_{-+}(\omega)\right]\quad,
\label{chi}
\end{equation}
which quantifies the probability of changing the spin-state by $\hbar$, and is directly related to the spectra measurable with ISTS measurements when probing spin excitations~\cite{Lounis2010,Lounis:2014in}.
This quantity is depicted in Fig.~\ref{spin-excitation}a for all the adatoms as function of the energy. 
Each curve is characterized by two quantities: the excitation energy and the width of the resonance peak.
The former is directly related to the MAE and, consequently, to the strength of the spin-orbit interaction~\cite{dosSantosDias:2015bh}.
The latter is shaped by the density of electron-hole excitations of opposite spin and thus proportional to the strength of the hybridization between the electronic states of the adatom with those of the substrate~\cite{Lounis:2015ho}. 

Since 4$d$ adatoms have generally larger MAE (as discussed in Sec.~\ref{groundstate}), their resonant energies are larger than those of 3$d$ adatoms. 
In particular, the largest resonance energy is observed for Mo, with a value of \SI{9.40}{\milli\electronvolt}, followed by Tc and Ru, with peaks located at \SI{5.83}{\milli\electronvolt} and \SI{2.71}{\milli\electronvolt}, respectively. These results correlate with the trend of the MAE, which opens the excitation gap, shown in Table~\ref{table2}.
The spin excitation energies related to Fe and Co are located at $\SI{0.22}{\milli\electronvolt}$ and $\SI{0.64}{\milli\electronvolt}$, respectively, while Mn exhibits a very small resonance energy of $\SI{0.02}{\milli\electronvolt}$.
The latter is in line with the weak MAE of Mn (see Table~\ref{table2} and Fig.~\ref{fig33}).


Next, we analyze the lifetime of the transversal spin excitations, which is a quantity of prime importance from the application point of view~\cite{loth_measurement_2010,baumann_electron_2015,delgado_spin_2017,IbanezAzpiroz:2017gc}.
The lifetime $\tau$ is defined as the inverse of the full-width at half maximum of the spin excitation resonance $\Gamma$, i.e., $\tau=\frac{2\hbar}{\Gamma}$~\cite{IbanezAzpiroz:2017gc, dosSantosDias:2015bh}. 
In a phenomenological description given by Eq.~\eqref{LLG_eq}, $\Gamma$ is proportional to the Gilbert damping $\alpha$~\cite{Guimaraes:2019bs}. 
By fitting the curves in Fig.~\ref{spin-excitation} to the susceptibility expressions obtained from Eq.~\eqref{LLG_eq}, we obtain the values for $\alpha$ and $\tau$ given in Table~\ref{table2}.
Among the 3$d$ elements, the lifetime $\tau$ is significantly smaller for Fe ($\SI{0.2}{\pico\second}$) as compared to Mn ($\SI{14}{\nano\second}$) and Co ($\SI{192}{\pico\second}$).
The low values of $\tau$ for Fe can be tied back to the very large value of the Gilbert damping (as shown in Table~\ref{table2}), which, in turn, arises from the details of the electronic structure around the Fermi level $\varepsilon_\text{F}$. 
Indeed, the introduction of impurities into graphene may induce impurity states in the vicinity of $\varepsilon_\text{F}$~\cite{Wehling:2007hi}, increasing the spectral occupation in this range (see Fig.~\ref{ldos}). 
In the particular case of Fe, the impurity state in the majority-spin channel occurs exactly at the Fermi energy. 
Due to the large $d$-state peak located also at $\varepsilon_\text{F}$ in the minority-spin channel, the Gilbert damping $\alpha$ is strongly enhanced, as it is proportional 
to $n_{\uparrow}(\varepsilon_\text{F})\,n_{\uparrow}(\varepsilon_\text{F})$~\cite{Lounis:2015ho}, resulting in the overdamped excitation spectra seen in Fig~\ref{spin-excitation}a.
For 4$d$ impurities; the lifetime is large for Tc with a value of $\SI{34}{\pico\second}$, while $\tau$ is $\SI{9.3}{\pico\second}$ and $\SI{1.7}{\pico\second}$ for Ru and Mo, respectively.

\subsection{Zero-point spin-fluctuations}

The quantum mechanical system composed by the electrons spins can also fluctuate spontaneously, even without any applied external field.
These are the ZPSF, which can also alter the magnetic stability of the adatoms through the modification of the MAE barrier~\cite{IbanezAzpiroz:2016fa}. 
The magnitude of the ZPSF, $\xi _{\perp }^{2}$, can be accessed via the fluctuation-dissipation theorem\cite{Kubo:2002dq},
\begin{equation}
\xi_{\perp }^{2} = -\frac{1}{\pi }\int_{0}^{+\infty }\hspace{-5mm} d\omega\, \text{Im}\,\chi
_{\bot }\left( \omega \right)\quad.
\label{sf_fluct}
\end{equation}
Note that the spontaneous fluctuations are defined in terms of the density of spin excitations given by the imaginary part of $\chi_{+-}$ (see Eq.~\ref{chi}).
It is obtained by integrating the spectra displayed in Fig.~\ref{spin-excitation}a over all positive energies.
Interestingly, $\xi_{\perp }^{2}$ is highly dependent on the position of spin-excitation peak and its linewidth~\cite{IbanezAzpiroz:2016fa}.
The calculated magnitude of the ZPSF, $\xi_\perp$ (in units of $\mu_\text{B}$), is displayed in Fig.~\ref{spin-excitation}b, together with the spin magnetic moment obtained for each impurity.
In Table~\ref{table2}, we list the ratio between the ZPSF and the spin magnetic moment, $\xi_{\perp}/{{M}_\text{s}}$.
Surprisingly, our \textit{ab initio} calculations reveal that the magnitude of the fluctuations for Co and Tc are larger than the spin magnetic moment itself, while being nearly as big for Mo and Ru (see Table~\ref{table2}).
In the case of Co, the high fluctuations are driven by the very large excitation amplitude, while Tc presents low amplitude but in a very large range of energies.
Mn and Fe are more protected against the fluctuations, with ${\xi _{\perp }}/{{M}_\text{s}}\simeq 20\%$.
The source of low fluctuations, however, is very different: while Mn presents a sharp excitation peak --- meaning that it is difficult to move the spin moment for most of the enegies ---, Fe has a very broad excitation due to its hitherto discussed high damping --- and even though the peak is relatively broad, the amplitude of excitation is fairly low.

\begin{figure}[tbp]
\includegraphics[width=1.0\columnwidth]{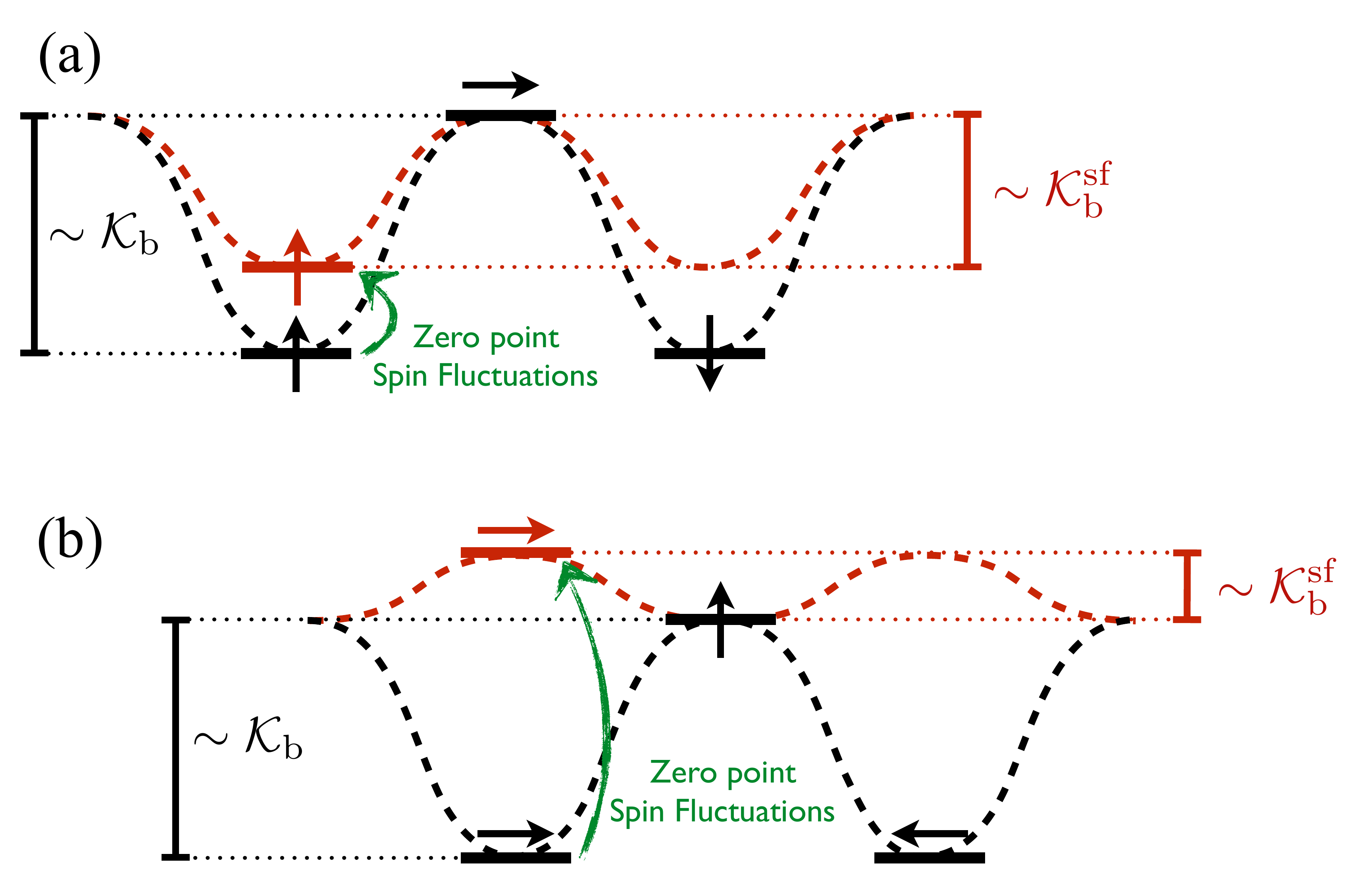}
\caption{Renormalization of the magnetic anisotropy energy due to the zero-point spin-fluctuations (ZPSF). (a) When the ZPSF are relatively small, the barrier is effectively reduced, as for example in the case of Fe. (b) When the fluctuations are larger then the magnetic moment ($\xi>M_\text{s}$), the easy axis is switched and magnetic bistability can be induced, as for Tc.}
\label{renormalized_barrier}
\end{figure}

The fluctuations affect the stability of the system by modifying the effective energy barrier between different spin orientations.
As the system fluctuates, the ground state is not in the minimum value of energy, as in a quantum harmonic oscillator.
The effective barrier that a system must overcome to change between states is then reduced.
This is quantified by the renormalized MAE, $\mathcal{K}^\text{sf}$, given by~\cite{IbanezAzpiroz:2016fa}
\begin{equation}
\mathcal{K}^\text{sf} =\mathcal{K}\left( 1-\frac{3\xi _{\perp
}^{2}}{{M_\text{s}}^{2}+2\xi _{\perp }^{2}}\right)\quad,
\label{MAE_renorm}
\end{equation}%
where $\mathcal{K}$ denotes the bare MAE. 
Eq.~\eqref{MAE_renorm} shows in clear fashion how the transversal fluctuations alter the height of the anisotropy barrier: larger values of $\xi_\perp$ results in a lowering of the MAE. 
A comparison between the bare anisotropies obtained from the band energy, $\mathcal{K}_\text{b}$, and from the susceptibility, $\mathcal{K}_\text{susc}$, as well as their renormalized counterparts are shown in Fig.~\ref{fig33}. 
3$d$ and 4$d$ impurities display rather contrasting behaviours, since the MAE for 3$d$ impurities remains nearly unaffected by the fluctuations, which is due to the fact that the fluctuations are relatively small for Mn and Fe. 
This change is represented in Fig~\ref{renormalized_barrier}a, where the original barrier defined by $\mathcal{K}_\text{b}$ is decreased to $\mathcal{K}^\text{sf}_\text{b}$.
The MAE of Co is negligible and is very weakly altered; we note in passing that the very weak anisotropy has been also observed experimentally for Co trimers on graphene~\cite{Wang:2012id}. 
For the 4$d$ elements, the fluctuations are substantial as they lead to a considerable reduction of the MAE barrier in the case of Mo and Ru. 
Interestingly, the most extreme scenario occurs for Tc impurities displaying the largest spin-fluctuations amplitude, where the renormalization switches the preferable moment orientation from in-plane to out-of-plane. 
This case is illustrated in Fig.~\ref{renormalized_barrier}b.
Counterintuitively, ZPSF promote an atom, being initially not interesting, to the pool of candidates for magnetic bistability. 
Among all the elements investigated so-far, it is the first time that ZPSF is not detrimental in the race of establishing single magnetic moments as stable entities for information technology.

\section{Conclusion}
\label{sec:conclusions}

In this work we have investigated the magnetic stability of single 3$d$ transition metal adatoms (Mn, Fe and Co), as well as 4$d$ adatoms (Mo, Tc and Ru), adsorbed on graphene by using first-principles DFT calculations based on the KKR approach. 
We have performed a systematic analysis of several groundstate and dynamical properties including the spin-excitation energy, the spin lifetime, the spin-fluctuation spectra and the ZPSF-renormalized MAE. 
Our static DFT calculations exhibit spin excitations of the adatoms with a large range of lifetimes, from fractions of picoseconds for Fe to nanoseconds for Mn.
Those values are highly influenced by the impurity states close to the Fermi energy, and indicate that the lifetime can be controlled with the application of electric gates.
The obtained ground-state results  establish perpendicular magnetic anisotropies for Fe and Co, whereas the rest of adatoms show an in-plane easy-axis. 
Nevertheless, zero-point spin-fluctuations can severely alter the static picture; apart from producing an overall decrease of the magnitude of the MAE, in extreme cases like Co and Tc adatoms, fluctuations can even reverse the preferred spin orientation.
This unexpected result for Tc demonstrates how the ZPSF can modify the energy landscape to promote bistability in an otherwise in-plane geometry.
Our analysis indicates that Fe is the best candidate for potential technological applications; apart from showing a perpendicular easy-axis, it is particularly robust against fluctuations.
Altogether, we believe our investigation has shed light into the dynamical magnetic properties of this fascinating material when coated with transition metal single adatoms.

\section*{Acknowledgements}
L. B. Drissi acknowledges the ``Acad\'{e}mie Hassan II des Sciences et Techniques-Morocco'' for financial support and the Alexander von Humboldt Foundation for financial support via the Georg Forster Research Fellowship for experienced scientists (Ref 3.4 - MAR - 1202992).               S. Sadki and L. B. Drissi 
acknowledge Peter Gr\"{u}nberg Institut and Institute for Advanced
Simulation, Forschungszentrum J\"{u}lich and JARA, J\"{u}lich, Germany for providing various 
facilities. J. Bouaziz, F. S. M. Guimar\~aes and S. Lounis acknowledge support from the European Research
Council (ERC) under the European Union's Horizon 2020 research and
innovation programme (ERC-consolidator Grant No. 681405 DYNASORE. S. Sadki, L. B.
Drissi and S. Lounis express their gratitude to the Arab German Young Academy of Sciences
and Humanities (AGYA) sponsored by the Federal Ministry of Education and
Research (BMBF) for funding.
J. Iba\~nez-Azpiroz received funding from the 
European Union's Horizon 2020 research and innovation
programme under the Marie Sklodowska-Curie grant agreement No 839237.

\bibliography{bibliography.bib}

\end{document}